# Gauged Wess–Zumino Terms and Equivariant Cohomology

José M Figueroa-O'Farrill and Sonia Stanciu [1,2]

*Department of Physics, Queen Mary and Westfield College, Mile End Road, London E1 4NS, UK*

**Abstract:** We summarize some results obtained on the problem of gauging the Wess–Zumino term of a $d$-dimensional bosonic $\sigma$-model. We show that gauged WZ-like terms are in one-to-one correspondence with equivariant cocycles of the target space. By the same token, the obstructions to gauging a WZ term can be understood in terms of the equivariant cohomology of the target space and this allows us to use topological tools to derive some a priori vanishing theorems guaranteeing the absence of obstructions for a large class of target spaces and symmetry groups in the physically interesting dimensions $d \leq 4$.

## 1. Introduction

The purpose of this note is to summarize some results on the gauging of bosonic $\sigma$-models with Wess–Zumino terms in arbitrary dimensions. These results are discussed in detail in [4]. The motivation for this work stems from the study of two-dimensional gauged WZW models based on Lie groups admitting a bi-invariant metric [5]. A bi-invariant metric restricts to an ad-invariant metric on its Lie algebra. Lie algebras possessing ad-invariant metrics [12] are called self-dual and they include, apart from those Lie algebras which are direct sums of semisimple and abelian Lie algebras, also nonreductive examples that have recently attracted a lot of attention. Indeed, it has been recently realized [14] that WZW models defined on nonreductive Lie groups give rise to conformal field theories from which exact string backgrounds can be constructed. The nonperturbative conformal invariance of these models is guaranteed by the existence of a quantum Sugawara construction [11,13,3] associated to the affinization of any self-dual Lie algebra. Gauging these models is interesting

---

because one can obtain more exact string backgrounds and because of the role it plays in string duality. However it is well-known that it is not possible to gauge all symmetry groups; but only those so-called "anomaly-free". It is the understanding of these obstructions that motivates the present work.

The two-dimensional WZW model is perhaps the simplest example of the obstructions we have in mind. Let $G$ be a Lie group with self-dual Lie algebra $\mathfrak{g}$, and let $\langle -, - \rangle$ denote the invariant metric on $\mathfrak{g}$. Since the metric on $G$ is bi-invariant, the isometry group of the WZW model with target $G$ is $G \times G$. We would like to know when can a subgroup $H \subset G \times G$ be gauged. The Lie algebra embedding $\mathfrak{h} \subset \mathfrak{g} \times \mathfrak{g}$ is characterized by two homomorphisms $\ell, r : \mathfrak{h} \to \mathfrak{g}$, so that if $X$ is an element of $\mathfrak{h}$, its image in $\mathfrak{g} \times \mathfrak{g}$ is $(\ell(X), r(X))$. Then the condition for $H$ to be gauged is simply that for all $X, Y \in \mathfrak{h}$, $\langle \ell(X), \ell(Y) \rangle = \langle r(X), r(Y) \rangle$. In other words, if we let $\langle -, - \rangle_\ell$ (resp. $\langle -, - \rangle_r$) denote the bilinear form induced on $\mathfrak{h}$ by pulling back $\langle -, - \rangle$ via $\ell$ (resp. $r$); then the condition is that they should agree:

$$\langle -, - \rangle_\ell = \langle -, - \rangle_r . \qquad (1)$$

Among the self-dual Lie algebras, the nonreductive ones have the interesting property that they have generically more subalgebras such that (1) is satisfied. In fact this property underpins the mechanism exhibited in [14] by which the beta function of the WZW model does not admit further quantum corrections beyond one-loop—the underlying reason for both phenomena being the existence of "large" isotropic subalgebras.

This note summarizes the result of our attempt to understand the precise nature of (1). However our setting will be more general. We set out to investigate the nature of the obstructions to gauging a $d$-dimensional bosonic $\sigma$-model with a Wess–Zumino term. These obstructions have been known for some time. The case $d = 1$ was investigated in [15] (but see also [2]), the case $d = 2$ is to be found in [7] and [9]; whereas the general case has been studied in [8]. After this work was finished we were made aware of the fact that the relation between gauging the WZ term and equivariant cohomology was already implicit in [15] for the case $d = 1$ and explicit in the appendix of [16] for the case $d = 2$.

## 2. Gauging $\sigma$-models

Here we set the notation. We let $\Sigma$ be the spacetime of the $\sigma$-model. It is taken to be a $d$-dimensional (pseudo)riemannian connected, oriented manifold and we let $\star$ denote the Hodge star associated to the metric and choice of orientation. The target space is taken to be any (pseudo)riemannian manifold

$(M, g)$. The action of the $\sigma$-model is a functional on maps $\varphi : \Sigma \to M$, and is given by

$$I_\sigma[\varphi] = \tfrac{1}{2} \int_\Sigma \|d\varphi\|^2 = \tfrac{1}{2} \int_\Sigma g_{ij}(\varphi) d\varphi^i \wedge \star d\varphi^j \ . \qquad (2)$$

Suppose now that a connected Lie group $G$ acts on $M$ via isometries, and let $\mathfrak{g}$ denote its Lie algebra. Fix a basis $\langle X_a \rangle$ for $\mathfrak{g}$ and let $\xi_a$ denote the corresponding Killing vectors on $M$. If $\lambda^a$ are constants, then the action (2) is invariant under

$$\delta_\lambda \varphi^i = \lambda^a \xi_a^i(\varphi) \ . \qquad (3)$$

If the $\lambda^a$ are functions on $\Sigma$, then the action $I_\sigma[\varphi]$ can still be made invariant under (3) provided that we minimally couple it to gauge fields $A = A^a X_a$. This is achieved by introducing the exterior covariant derivative

$$\nabla \varphi^i = d\varphi^i - A^a \xi_a^i(\varphi) \qquad (4)$$

and defining

$$I_G[\varphi, A] = \tfrac{1}{2} \int_\Sigma \|\nabla \varphi\|^2 = \tfrac{1}{2} \int_\Sigma g_{ij}(\varphi) \nabla \varphi^i \wedge \star \nabla \varphi^j \ . \qquad (5)$$

On the other hand, we will see that things are not so simple when trying to gauge a Wess–Zumino term.

Indeed, let $B$ be a $(d+1)$-dimensional manifold with boundary $\partial B = \Sigma$ and let $\overline{\varphi} : B \to M$ denote an extension of $\varphi$ to $B$. We assume tacitly that all obstructions standing in the way of such an extension are overcome. If $\omega \in \Omega^{d+1}(M)$ is a closed form, then the variation of the Wess–Zumino term

$$I_{\mathrm{WZ}}[\overline{\varphi}] = \int_B \overline{\varphi}^* \omega \qquad (6)$$

is a boundary term and hence depends only on $\varphi$. If $\omega$ is in addition $G$-invariant, $I_{\mathrm{WZ}}$ will also be invariant under constant gauge transformations $\delta_\lambda \overline{\varphi} = \lambda^a \xi_a^i(\overline{\varphi})$. Contrary to the action (2), minimal coupling does not suffice to render (6) gauge-invariant; for the variation of the minimally-coupled action will depend on the extension $\overline{\varphi}$. In fact, a gauged WZ term is by definition a gauge-invariant action such that its variation depends only on $\varphi$ and fields defined on $\Sigma$. It has been known for some time that this is not always possible and that one must overcome a series of obstructions. We will now re-interpret these obstructions in terms of the equivariant cohomology of the target space and use this characterization to guarantee the absence of obstructions for a large class of target spaces and symmetry groups in the physically interesting dimensions $d \leq 4$.

## 3. Gauge Invariance and Equivariant Cohomology

We will actually obtain "universal" obstructions. That is, we will work with forms on $M$ and not with their pull-back to $B$ (or $\Sigma$). Similarly our gauge fields $A^a X_a$ will have no direct geometric meaning on $\Sigma$, but will be abstract generators of a differential algebra (DGA). This DGA is called the Weil algebra and is denoted by $W(\mathfrak{g})$. It is defined as follows: it is freely generated by "forms" $A^a$ and $F^a$ of degrees 1 and 2 respectively, subject to the condition

$$F^a = dA^a + \tfrac{1}{2} f_{bc}{}^a A^b A^c \,, \tag{7}$$

and where $d$ is extended to all of $W(\mathfrak{g})$ as an antiderivation obeying $d^2 = 0$. The Weil algebra $W(\mathfrak{g})$ is a universal model for the DGA generated by the connection 1-form and the curvature 2-form of a principal $G$-bundle, in the sense that there is a unique homomorphism of DGAs (the Weil homomorphism) sending $A^a$ and $F^a$ to the components of the connection 1-form and curvature 2-form respectively.

Under the gauge transformations $\delta_\lambda x^i = \lambda^a \xi^i_a(x)$ on $M$, where the $x^i$ are local coordinates and $\lambda^a$ need not be a constat, a differential form $\phi \in \Omega(M)$ transforms according to

$$\delta_\lambda \phi = d\lambda^a \imath_a \phi + \lambda^a \mathcal{L}_a \phi \,, \tag{8}$$

where $\imath_a$ is the contraction with the Killing vector $\xi_a$ and $\mathcal{L}_a \equiv [d, \imath_a]$ is the Lie derivative. Similarly, the transformation properties of the connection and curvature forms are given by

$$\begin{aligned} \delta_\lambda A^a &= d\lambda^a - f_{bc}{}^a \lambda^b A^c \\ \delta_\lambda F^a &= -f_{bc}{}^a \lambda^b F^c \,. \end{aligned} \tag{9}$$

We can write these transformations in a form resembling (8) as follows. We introduce an abstract contraction $\imath_a$ on the Weil algebra by $\imath_a A^b = \delta_a{}^b$ and $\imath_a F^b = 0$, and extending it to all of $W(\mathfrak{g})$ as an antiderivation. We then define an abstract Lie derivative by $\mathcal{L}_a \equiv [d, \imath_a]$. Then substituting $A^a$ and $F^a$ in turn for $\phi$ in (8) yields (9).

In other words, we can consider the DGA $W(\mathfrak{g}) \otimes \Omega(M)$ with (anti)derivations $d$, $\imath_a$ and $\mathcal{L}_a$ defined on each tensorand in the appropriate fashion. Then if $\phi \in W(\mathfrak{g}) \otimes \Omega(M)$ is any form, its gauge variation is given by (8). This immediately implies that $\phi$ is gauge-invariant if and only if $\imath_a \phi = 0$ and $\mathcal{L}_a \phi = 0$ for all $a$. The former condition says that $\phi$ is horizontal, whereas the latter says that it is invariant. Forms which are horizontal and invariant are called basic or equivariant and they form a subalgebra closed under $d$ (since $d$ and $\delta_\lambda$ commute) which is denoted $\Omega_G(M)$. Its cohomology $H_G(M)$ is called the (algebraic) $G$-equivariant cohomology of $M$ (see, for instance, [1]).

## 4. The Obstruction

Thus we see that the Wess–Zumino term can be gauged precisely when $\omega \in \Omega^{d+1}(M)$ admits an equivariant closed extension $\omega^{\#} \in \Omega^{d+1}_G(M)$. In other words, when we can add $(A, F)$-dependent terms to $\omega$ in such a way that the resulting form is both gauge-invariant and closed. We now analyze this condition homologically. If $\phi \in \Omega_G(M)$, we can put $A{=}F{=}0$ and the result is a $G$-invariant form on $M$. If we let $\Omega(M)^G$ denote the subcomplex of $G$-invariant forms on $M$, setting $A{=}F{=}0$ yields a map

$$\Phi : \Omega_G(M) \longrightarrow \Omega(M)^G , \qquad (10)$$

which commutes with the action of $d$. Minimal coupling gives a partial inverse to this map. In fact, if we let $\kappa : \Omega(M)^G \to \Omega_G(M)$ denote minimal coupling, we find that $\Phi(\kappa(\omega)) = \omega$. This means that $\Phi$ is surjective. Let us denote its kernel by $\Omega_\Phi(M)$. These are the gauge-invariant forms whose $(A, F)$-independent part vanishes. Since $\Phi$ commutes with $d$, $\Omega_\Phi(M)$ is closed under $d$. These remarks mean that we have a short exact sequence of complexes

$$0 \longrightarrow \Omega_\Phi(M) \longrightarrow \Omega_G(M) \xrightarrow{\Phi} \Omega(M)^G \longrightarrow 0 . \qquad (11)$$

This gives rise to a long exact sequence in cohomology, of which the relevant bit is given by

$$\cdots \longrightarrow H^{d+1}_G(M) \xrightarrow{\Phi_*} H^{d+1}(M)^G \xrightarrow{d_*} H^{d+2}_\Phi(M) \longrightarrow \cdots \qquad (12)$$

This construction gives rise to the following:

**Theorem 1** *The obstruction to gauging $\omega \in \Omega^{d+1}(M)^G$ is the class $d_*[\omega] \in H^{d+2}_\Phi(M)$.*

**Proof.** By exactness of (12), $d_*[\omega] = 0$ precisely when $[\omega] = \Phi_*[\widehat{\omega}]$ for some equivariant cocycle $\widehat{\omega}$. That is, if there is an equivariant cocycle $\widehat{\omega}$ such that $\Phi(\widehat{\omega}) = \omega + d\theta$, where $\theta$ is some invariant form on $M$. If we then let $\kappa(\theta)$ be the equivariant form obtained from $\theta$ by minimal coupling, then $\omega^{\#} \equiv \widehat{\omega} - d\kappa(\theta)$ is the desired closed equivariant extension of $\omega$: for $\omega^{\#}$ is clearly closed and moreover $\Phi(\omega^{\#}) = \omega$. Conversely, if $\omega$ has an equivariant closed extension $\omega^{\#}$, then $[\omega] = \Phi_*[\omega^{\#}]$, hence $d_*[\omega] = 0$. □

Notice that if $[\omega] = 0$, then the obstruction vanishes. This is nothing but the fact that if $\omega = d\theta$ with $\theta$ invariant, then we can simply gauge the action by minimal coupling. In other words, we let $\kappa(\theta)$ denote the minimally coupled $\kappa$ and we define the gauged action as the integral over $\Sigma$ of the pull-back of $\kappa(\theta)$, where pulling back a form in $W(\mathfrak{g}) \otimes \Omega(M)$, means applying the geometric pull-back on the second tensorand and the Weil homomorphism on the first. The resulting action is gauge-invariant but it is no longer a topological term.

## 5. The Cartan Model and the Obstructions of Hull and Spence

In any local gauge-invariant quantity, the gauge fields $A^a$ appear only via the covariant derivative. It therefore makes sense that we should be able to dispense with them in computing equivariant cohomology. This is indeed the case and $H_G(M)$ can be computed from another complex involving only the curvature $F^a$ and the forms on $M$. This is called the Cartan model for equivariant cohomology (see, for instance, [10]). The space of forms in the Cartan model is given by

$$\mathcal{C} = (\mathfrak{S}\mathfrak{g}^* \otimes \Omega(M))^G \; ; \tag{13}$$

or in other words, finite linear combinations of $G$-invariant monomials of the form $F^a \cdots F^b \omega_{a \cdots b}$, where $\omega_{a \cdots b}$ are forms on $M$. The invariance condition says that

$$\mathcal{L}_a \omega_{b \cdots c} = f_{ab}{}^d \omega_{d \cdots c} + \cdots + f_{ac}{}^d \omega_{b \cdots d} \; . \tag{14}$$

The differential $D$ of the Cartan model is defined by $DF^a = 0$ and

$$D\omega = d\omega - F^a \imath_a \omega \qquad \text{for } \omega \in \Omega(M) \; , \tag{15}$$

and extended to all of $\mathcal{C}$ as an antiderivation. Notice that $\mathcal{C}$ is indeed closed under $D$ because the differential is $G$-invariant; and also that indeed $D^2 = 0$. The grading on $\mathcal{C}$ is the one inherited from $W(\mathfrak{g}) \otimes \Omega(M)$, so that if $\omega_{a_1 \cdots a_p}$ are $r$-forms on $M$, the monomial $F^{a_1} \cdots F^{a_p} \omega_{a_1 \cdots a_p}$ has degree $2p + r$. It is then clear that $D$ has degree 1.

The obstructions found by Hull and Spence in [8] are easy to recover in the Cartan model. We first notice that the Cartan model $(\mathcal{C}, D)$ restricts nicely to a model $(\mathcal{C}_\Phi, D)$ for $H_\Phi(M)$. Indeed, $\mathcal{C}_\Phi$ is the subalgebra of $\mathcal{C}$ without $F$-independent monomials. It is clearly closed under $D$ and moreover it is easy to prove that its $D$-cohomology is precisely $H_\Phi(M)$.

We now apply the theorem. The obstruction is the class $d_*[\omega]$ in $H_\Phi^{d+2}(M)$. In the Cartan model, $d_*[\omega]$ is represented by the cocycle $D\omega = -F^a \imath_a \omega$. The obstruction class vanishes if there exists $\Delta \in \mathcal{C}_\Phi^{d+1}$, such that

$$D\Delta = -F^a \imath_a \omega \; . \tag{16}$$

Expanding $\Delta$ as follows

$$\Delta = F^a \Delta_a + \tfrac{1}{2} F^a F^b \Delta_{ab} + \cdots \tag{17}$$

where each $\Delta_{a\cdots b}$ obeys the invariance condition (14), we find that equation (16) becomes the set of conditions

$$\begin{aligned}
-\imath_a \omega &= d\Delta_a \\
\imath_a \Delta_b + \imath_b \Delta_a &= d\Delta_{ab} \\
&\vdots \\
\sum_{i=1}^{p} \imath_{a_i} \Delta_{a_1 \cdots \widehat{a_i} \cdots a_p} &= d\Delta_{a_1 \cdots a_p}\ , \\
&\vdots
\end{aligned}$$

where the $\widehat{\phantom{a}}$ over an index denotes its omission. These are precisely the obstructions discovered by Hull and Spence in [8] and rediscovered by Wu in [17].

If such a $\Delta$ exists, the equivariant closed extension $\omega^\#$ of $\omega$ is obtained by re-introducing the gauge fields in $\omega - \Delta$ via minimal coupling. That is, $\omega^\# = \kappa(\omega - \Delta)$. Notice that even if $\Delta$ exists, it need not be unique. Clearly any other $\Delta' \in \mathcal{C}_\Phi^{d+1}$ would also cancel the obstruction provided the difference $T = \Delta - \Delta'$ were $D$-closed. This means that we can add a term

$$\int_B \overline{\varphi}^* \kappa(T) \tag{18}$$

to the action, where $\kappa(T)$ is the minimally coupled $T$. If $T$ is $D$-exact, say $T = DU$, then $\kappa(T) = d\kappa(U)$ and (18) becomes

$$\int_\Sigma \varphi^* \kappa(U) \tag{19}$$

which is not longer a topological term. On the other hand, if $T$ defines a nontrivial class in $H_\Phi^{d+1}(M)$, then the action (18) defines a new topological term for the $\sigma$-model. Moreover this topological term has the property that it is not obtained by gauging a Wess–Zumino term, since $T = 0$ when $A = F = 0$. Topological terms of this kind were first considered by Hull, Roček, and de Wit in [6]. The above terms include and generalize the ones in [6], because whereas the terms in [6] exist only for nonsemisimple $G$ and for even dimensions $d = 2k$, it can be shown that $H_\Phi^{d+1}(M)$ may be nontrivial in any dimension and for arbitrary $G$. Indeed, the triviality of $H_\Phi^{d+1}(M)$ would imply that there are no obstructions to gauging the Wess–Zumino term of a $(d-1)$-dimensional $\sigma$-model with symmetry $G$, which can be shown not to be the case generically even if $G$ were semisimple.

## 6. Vanishing Theorems

Having identified the cohomology in which the obstruction class lives, we can now try to analyze in what circumstances the relevant cohomology spaces vanish. This would immediately imply the absence of obstructions to gauge any Wess–Zumino term. Of course, this vanishing of cohomology is generally too strong a condition and in many cases it still may be that the cohomology is nontrivial but the obstruction class is. This happens for instance in the WZW model with the standard Wess–Zumino term. A priori we would expect—as the dimension increases—an equally increasing number of obstructions that must be satisfied; but in fact there is only one obstruction which can be seen to live in the Casimir ring of the symmetry algebra. For the $d=2$ WZW model, this is exemplified by equation (1).

The vanishing theorems obtained in this section are valid when $G$ is a compact Lie group. This condition on the topology of the Lie group should not be essential given the algebraic nature of the obstructions, and in [4] it is argued that the compactness assumption can be dropped without altering the results below, provided that the target space possesses a $G$-equivariant finitely generated model (in the sense of rational homotopy theory). We feel that the existence of these equivariant models—which to the best of our knowledge has not been established—is true and we have tried to persuade the reader by constructing a few simple examples in [4].

If $G$ is compact, the algebraic equivariant cohomology of $M$ defined in Section 3 agrees with the real singular cohomology $H(M_G; \mathbb{R})$ of the homotopy quotient $M_G \equiv EG \times_G M$, where $EG$ is the total space of the universal $G$-bundle $EG \to BG$. It can be shown that if the action of $G$ on $M$ is free, then $M_G$ has the same homotopy type as the quotient $M/G$, whence $H_G(M) \cong H(M/G)$. In any case, it is easy to show that $M_G$ is a fiber bundle over $BG$ with typical fiber $M$, and hence its real singular cohomology can be approximated by the Leray spectral sequence. This spectral sequence adapts itself to a spectral sequence converging to $H_\Phi(M)$, and we have the following [4]

**Theorem 2** *There exists a spectral sequence $(E_r^{\bullet,\bullet}, d_r)$ with*

$$d_r : E_r^{p,q} \longrightarrow E_r^{p+r,q-r+1} \tag{20}$$

*with $E_r^n = \bigoplus_{\substack{p \geq 2, q \geq 0 \\ p+q=n}} E_r^{p,q}$,*

$$E_2^{p,q} = H^p(BG; \mathbb{R}) \otimes H^q(M; \mathbb{R}) \cong \begin{cases} \left(\mathfrak{S}^{p/2}\mathfrak{g}^*\right)^G \otimes H^q(M) & \text{for } p \text{ even;} \\ 0 & \text{for } p \text{ odd.} \end{cases} \tag{21}$$

*and such that it converges to $H_\Phi^\bullet(M)$.*

It follows that if we find that $E_2^n = 0$ for some $M$ and $G$, then $H_\Phi^n(M) = 0$ also. As $d$ increases the vanishing of $E_2^{d+2}$ imposes more and more conditions on the topology of $M$ and the structure of $G$, but for the physically interesting dimensions ($d \leq 4$) we find conditions which are not so restrictive. If $\mathfrak{g}$ is semisimple, notice that $(\mathfrak{g}^*)^G = 0$, so that the first $E_2$'s are as follows:

$$E_2^4 = \left(\mathfrak{S}^2 \mathfrak{g}^*\right)^G \otimes H^0(M)$$
$$E_2^5 = \left(\mathfrak{S}^2 \mathfrak{g}^*\right)^G \otimes H^1(M)$$
$$E_2^6 = \left(\left(\mathfrak{S}^3 \mathfrak{g}^*\right)^G \otimes H^0(M)\right) \oplus \left(\left(\mathfrak{S}^2 \mathfrak{g}^*\right)^G \otimes H^2(M)\right) \ .$$

These are approximations to $H_\Phi^{d+2}(M)$ for $d=2, 3, 4$, which contain obstructions for $\sigma$-models in dimension $d$. Whatever the topology of $M$, $H^0(M)$ is never trivial, and regardless the structure of $G$, $(\mathfrak{S}^2 \mathfrak{g}^*)^G$ contains at least the Killing form. Therefore $E_2^4$ is generically nontrivial and we cannot discard the existence of an obstruction for a $d = 2$ $\sigma$-model. We know this to be the case for the WZW model as discussed in the introduction, where the obstruction $\langle -, -\rangle_\ell - \langle -, -\rangle_r$ defines an element in $E_2^4 \cong H_\Phi^4(M)$. For $d = 3$ things are different. Indeed, we see that if $H^1(M) = 0$, (e.g., $M$ simply-connected), there are no obstructions. Similarly if $H^2(M) = 0$ and $\mathfrak{g}$ has not cubic casimirs, there are no obstructions to gauging the WZ term in a four-dimensional $\sigma$-model. We can summarize these observations as follows:

**Corollary 3** *For any three-dimensional $\sigma$-model with WZ term whose target manifold $M$ obeys $H^1(M) = 0$ any compact semisimple group can be gauged. This holds, in particular, if $M$ is simply connected.*

**Corollary 4** *For any four-dimensional $\sigma$-model with WZ term whose target manifold $M$ obeys $H^2(M) = 0$, any compact semisimple group without a cubic casimir can be gauged. This is true, in particular, if the target manifold is a Lie group whose maximal compact subgroup has at most one $U(1)$ factor.*

## 7. Conclusion

In summary, we see that there is a one-to-one correspondence between topological terms for a gauged bosonic $\sigma$-model and equivariant cocycles. In particular, the obstruction to gauging a Wess–Zumino term can be understood as an obstruction to extend a given invariant form on $M$ to an equivariant form in $W(\mathfrak{g}) \otimes \Omega(M)$. This obstruction is measured cohomologically by $H_\Phi(M)$—the cohomology of a subcomplex of the equivariant forms on $M$, defined in Section 4. This cohomology theory seems to have more physical applications, since it is also responsible for the existence of new topological terms gener-alizing those of [6]. Finally we have seen that for compact semisimple groups we can derive vanishing theorems for $H_\Phi(M)$ which guarantee the absence of obstructions in physically interesting dimensions ($d \leq 4$) for a large class of target spaces. Spacetime constraints force the treatment to be sketchy in this announcement. For the full details as well as for other aspects of this topic which we could not cover here, the reader is referred to [4].

**Acknowledgement**

It is a pleasure to thank Chris Hull for drawing our attention to this beautiful topic and for many enlightening conversations. We have also benefited greatly by electronic conversations with Takashi Kimura who suffered through earlier versions of the results in this paper as we streamlined the presentation, and especially Jim Stasheff for his encouragement and his detailed comments on the draft version of [4]. We would like to thank Jaap Kalkman for sending us a copy of his thesis [10]; Siye Wu for sending us his paper [17] and for pointing to us the appendix of [16]; and George Papadopoulos for making us aware of [15]. One of us (SS) is grateful to the Department of Physics of Queen Mary and Westfield College, and Chris Hull in particular, for the opportunity to visit QMW this past academic year. Finally, we would like to acknowledge a contract of the European Commission Human Capital and Mobility Programme which partially funded this research.

This paper is archived as `hep-th/9407196`.